\documentclass[10pt,conference]{IEEEtran}
\usepackage{amsmath, balance}
\usepackage{tikz}
\usepackage{arydshln}
\usepackage{caption}
\usepackage{adjustbox}
\usepackage{verbatim}

\usepackage{xcolor}
\definecolor{light-gray}{gray}{0.95}

\definecolor{purple}{HTML}{AF72B0}

\begin{document}


\title{Model-Agnostic Syntactical Information for Pre-Trained Programming Language Models}

\author{\IEEEauthorblockN{Iman Saberi\IEEEauthorrefmark{1},
Fatemeh H. Fard\IEEEauthorrefmark{2}}
\IEEEauthorblockA{Department of Computer Science, Mathematics, Physics and Statistics\\
The University of British Columbia\\
Kelowna, Canada\\
Email: \IEEEauthorrefmark{1}iman.saberi@ubc.ca,
\IEEEauthorrefmark{2}fatemeh.fard@ubc.ca}}
\maketitle


\IEEEtitleabstractindextext{%
\begin{abstract}
Pre-trained Programming Language Models (PPLMs) achieved many recent states of the art results for many code-related software engineering tasks. 
Though some studies use data flow or propose tree-based models that utilize Abstract Syntax Tree (AST), most PPLMs do not fully utilize the rich syntactical information in source code. Still, the input is considered a sequence of tokens. There are two issues; the first is computational inefficiency due to the quadratic relationship between input length and attention complexity. Second, any syntactical information, when needed as an extra input to the current PPLMs, requires the model to be pre-trained from scratch, wasting all the computational resources already used for pre-training the current models. 
In this work, we propose \textit{Named Entity Recognition (NER) adapters}, lightweight modules that can be inserted into Transformer blocks to learn type information extracted from the AST. These adapters can be used with current PPLMs such as CodeBERT, GraphCodeBERT, and CodeT5. We train the NER adapters using a novel \textit{Token Type Classification objective function (TTC)}. 
We insert our proposed work in CodeBERT, building \textit{CodeBERTER}, and evaluate the performance on two tasks of code refinement and code summarization. 
CodeBERTER improves the accuracy of code refinement from 16.4 to 17.8 while using 20\% of training parameter budget compared to the fully fine-tuning approach, and the BLEU score of code summarization from 14.75 to 15.90 while reducing 77\% of training parameters compared to the fully fine-tuning approach.

\end{abstract}

}

\maketitle
\thispagestyle{plain}
\pagestyle{plain}
\IEEEdisplaynontitleabstractindextext

\IEEEpeerreviewmaketitle

\section{Introduction}
\label{section:introduction}
The success of Pre-trained Language Models (PLM) in Natural Language Processing (NLP) led to the emergence of pre-trained models in software engineering, such as CodeBERT \cite{feng2020codebert}, CuBERT \cite{kanade2019pre}, and CodeT5 \cite{wang2021codet5}. Pre-trained Programming Language Models (PPLMs) --PLMs pre-trained on programming languages-- learn a generic representation of code using an abstract objective function such as Mask Language Modeling (MLM) during the pretraining phase. These representations later are exploited for programming language-oriented downstream tasks such as code summarization \cite{gu2022assemble,ahmed2022learning,zhang2022survey}, code clone detection \cite{buch2019learning,nafi2019clcdsa} and code refinement \cite{guo2020graphcodebert,tufano2019empirical} in the fine-tuning phase.

Despite the advance of PPLMs, most of them \cite{kanade2019pre,lu2021codexglue,ahmad2021unified} exploit either the encoder or the decoder Transformer blocks \cite{vaswani2017attention}. These models treat a code snippet as a sequence of tokens like conventional techniques for input modeling in the natural language domain while ignoring the rich syntactical information of source code. Recently, several studies proposed representations with the structural information of source code in mind \cite{guo2020graphcodebert,wang2021codet5,jiang2021treebert}.

Jiang et al. proposed a tree-based PLM for programming language generation tasks \cite{jiang2021treebert}, which utilizes the Abstract Syntax Tree (AST) to provide structural information for pre-training a Transformer-based architecture \cite{vaswani2017attention}. Compared to the sequence-based approaches that consider code tokens as the model's input, in their approach, each code token is represented as a path from the root node to the terminal node corresponding to that code token in the AST. 
Though their approach feeds rich structural information to the model, the model's input length is multiplied by the average length of all paths in the AST. 
This scaling in the input length is not computationally efficient, according to Ainslie et al., who mentions that the computational and memory complexity of attention operation scales quadratically as the input length increases \cite{ainslie2020etc}.

Some other PPLMs provide type information to the existing models. GraphCodeBERT \cite{guo2020graphcodebert} considers the inherent structure of code and utilizes the dataflow extracted from AST corresponding to each input to provide semantic-level information as an additional input for the model; then pre-trained CodeBERT \cite{feng2020codebert} using this new information. Furthermore, authors of CodeT5 \cite{wang2021codet5} mentioned that developer-assigned identifiers consist of rich code semantics, so they proposed a novel identifier-aware pre-training task to notify the model whether a code token is an identifier or not. Both of these approaches require the model to be pre-trained from scratch. Even the new information provided by GraphCodeBERT, though this model is pre-trained on CodeBERT, requires pre-training from scratch, which is not time and computationally efficient.

Therefore, we cannot directly take advantage of the existing PPLMs such as CodeBERT \cite{feng2020codebert}, GraphCodeBERT \cite{guo2020graphcodebert}, and CodeT5 \cite{wang2021codet5} to propose a new information or input format. As these models are pre-trained based on a flattened sequence of code tokens \cite{wu2022survey}, proposing a new form of input representation requires all these models to be \textit{pre-trained from scratch}. 
These two issues, namely computational inefficiency of using syntactical information of source code (due to length increase) and requiring to pre-train the PPLMs from scratch when introducing new input, motivated us to propose an approach that addresses them. Thus, this work aims to provide \textit{syntactical embeddings of the source code} to the \textit{existing} pre-trained programming language models. 


In order to propose a parameter-efficient and model-agnostic approach for imposing syntactical information to the current PPLMs, we utilize \textit{adapters}. In NLP, an adapter is a lightweight component placed inside each Transformer block to modify its behavior \cite{houlsby2019parameter}. During adapter training, the weights of the pre-trained model are fixed, and the newly introduced adapter weights are trained on an objective function. Adapters are often used to adapt existing NLP models to work in a specific context (i.e., transfer learning) \cite{goel2022cross,pfeiffer2020mad} or to perform a particular task (i.e., quick fine-tuning) \cite{he2021effectiveness,le2021lightweight}. 
However, adapters have never been used to capture the syntactical information of source code nor to provide new information to the existing PLMs or PPLMs.



In this work, we propose \textbf{Named Entity Recognition (NER) Adapters}, lightweight modules inserted inside Transformer blocks whose aim is to learn type information extracted from the AST. These pluggable modules can be inserted into current PPLMs such as CodeBERT \cite{feng2020codebert}, GraphCodeBERT \cite{guo2020graphcodebert}, and CodeT5 \cite{wang2021codet5}. 
To train NER adapters, we pose the problem as a token classification task in which each input token is assigned a type extracted from the AST, and the NER adapter aims to detect the type of each token correctly. 

We insert NER adapters in CodeBERT \cite{feng2020codebert}, thus, developing a model that we refer to as \textbf{CodeBERTER}. We conduct our experiments on Java bug fixing data introduce in \cite{tufano2019empirical} for code refinement, and CodeSearchNet dataset \cite{husain2019codesearchnet}  for code summarization task. The main research question that we investigate here is: \textbf{Can NER adapters improve the performance of software engineering target tasks, namely, code refinement and code summarization, while using fewer trainable parameters? }



The results show that for code refinement, we improved 
the CodeBERT baseline accuracy from 16.4 to 17.8 with 23 Million trainable parameters, which is 80\% less than the total number of the baseline trainable parameters. We also applied NER adapters on code summarization, which improved the BLEU-4 score of two languages, including Ruby and Go, by $30\%$ and $29\%$, respectively, with 77\% fewer parameters compared to full fine-tuning.



Our main contributions are as follows:
\begin{itemize}
    \item Introduce NER adapters with a novel loss function, Token Type Classification Loss (TTC), to impose syntactical information to the existing PPLMs. We also publish the source code\footnote{https://github.com/ist1373/NER\_Adapters/}.
    
    \item Evaluate the performance of NER adapters on code refinement and code summarization.
    
    \item Improve the results of these tasks while training fewer parameters and being more computationally efficient.

\end{itemize}

The rest of the paper is organized as follows. In Section \ref{sec:background}, we provide an overview of important background information and then introduce NER adapters in Section \ref{sec:neradapter}. We provide the experiments setup and the details of our study in Section \ref{sec:expset}. Results and discussion are explained in Section \ref{sec:results}. Sections \ref{section:related-work} and \ref{sec:threat} are dedicated to the related works and threats to validity. Finally, we conclude the paper in Section \ref{section:conclusion}.

\section{Background}
\label{sec:background}
\subsection{Abstract Syntax Tree}

An AST is a tree-like representation of the syntax of a programming language. Each node in the tree represents a syntactic construct in the source code, such as a keyword, an operator, a variable, or a function. The structure of the tree reflects the structure of the program, with the root node representing the top-level construct and the leaf nodes representing the smallest, most basic constructs \cite{zhang2019novel}. 

ASTs can be used for a variety of purposes, including analyzing the structure of a program to check for syntax errors or enforce coding standards \cite{wendel2015thinking}, and extracting information about the program, such as variable and function definitions, to create documentation or generate code coverage reports \cite{lin2021improving}.

\subsection{Pre-trained Programming Language Models}
Pre-trained Programming Language Models (PPLMs) are deep neural language models trained on a large dataset of source code and learn to predict the next word or token in the code given the context of the previous words or tokens. These models later can be used to perform a variety of target tasks during the fine-tuning phase, such as code summarization \cite{wang2021codet5,hu2018summarizing}, code completion \cite{takerngsaksiri2022syntax}, code generation \cite{wang2021codet5}, and code refinement \cite{guo2020graphcodebert}.


\subsection{Adapters}

In natural language processing, adapters are lightweight components that can be used to adapt a language model to a specific task or dataset. Adapter-based training is an efficient and quick fine-tuning method that requires fewer parameters compared to traditional full fine-tuning. In comparison to simply adding a new head on top of a pre-trained language model (PLM), adapters offer superior performance due to their ability to integrate into the internal structure of the PLM and influence the network's internal embeddings.

Adapters are employed in several areas, including (1) fine-tuning, which is the process of continuing to train a machine learning model on a new dataset, using the parameters and weights learned from the original training dataset as a starting point and (2) domain adaptation is another area that adapters are used, which is the process of adapting a machine learning model to a new domain, or a specific subject area or context. This can be done by continuing to train the model on a dataset that is representative of the new domain or by adding domain-specific information to the model, such as domain-specific word embeddings or features.

\subsection{Language Adapters}
Language adapters' aim is to learn language-specific transformations \cite{pfeiffer2020mad}. They are trained on unlabeled data using an abstract objective function such as mask language modeling (MLM). They consist of a down-projection and an up-projection at each layer, with a residual connection between them. The down-projection, D, is a matrix with dimensions h x d, where h is the hidden size of the transformer model and d is the dimension of the adapter. The up-projection, U, is a matrix with dimensions d x h. The language adapter at each layer takes in the Transformer hidden state, $h_l$, and the residual, $r_l$, and applies the down-projection and up-projection to them, with a ReLU activation function. The output of the language adapter is then added to the residual connection. 

\begin{equation}
LanguageAdapter_l(h_l,r_l)=U_l(ReLU(D_l(h_l)))+r_l
\end{equation}

\subsection{AdapterFusion}
AdapterFusion is a method of combining the knowledge from different language adapters in order to improve performance on downstream tasks such as code summarization \cite{pfeiffer2020adapterfusion}. Given a set of $N$ language adapters, adapter fusion involves taking the weighted sum of the outputs of these adapters while the weights of the pre-trained model and the language adapters are fixed. AdapterFusion consists of key, value, and query matrices at each layer. The goal of AdapterFusion is to find the optimal combination of language adapters by minimizing a loss function using a target task. This allows for the extraction and composition of knowledge from different language adapters in order to improve performance on downstream tasks.

\begin{equation}
\Phi= \textrm{argmin } L(D;\Theta,\theta_1,...,\theta_N)
\end{equation}

where $\Phi$ consists of $Key_l$, $Value_l$ and $Query_l$ metrics at each layer $l$, as shown in Fig. \ref{fig:nerarch}. At each transformer block, the output of the feed-forward sub-layer is taken to be as the $Query$, and the output of each language adapter is used for both $Key$ and $Value$ vectors.



\section{NER Adapters}
\label{sec:neradapter}

The NER adapter aims to incorporate token-type information into the network, as token types can provide valuable syntactical information to a model. According to research in \cite{sharma2022exploratory}, different token types can have varying levels of importance on a pre-trained language model; for example, \textit{identifiers} are more significant than other types in terms of the amount of attention from the model and their learned representations. 
To train the model on code token types, we first need to identify each token type in our dataset. We then introduce a new variation of adapters, NER adapters, which we plug into the PPLM and train them on a token-type classification task. 
The NER adapter is expected to predict the type of each token when the adapter-training phase is done. 
We discuss the details of NER adapters in the three sub-sections below: structure, objective function, and training phase.
In the final subsection, we explain how NER adapter is placed in the transformer blocks of a PPLM.

\subsection{Structure}

We use the same architecture of language adapters for NER adapters, so they consist of down- and up-sampling feedforward blocks combined with residual connections:

\begin{equation}
NER Adapter_l(h_l,r_l)=U_l(ReLU(D_l(h_l)))+r_l
\end{equation}

where $h_l$ and $r_l$ are the hidden state and residuals at layer $l$, respectively.

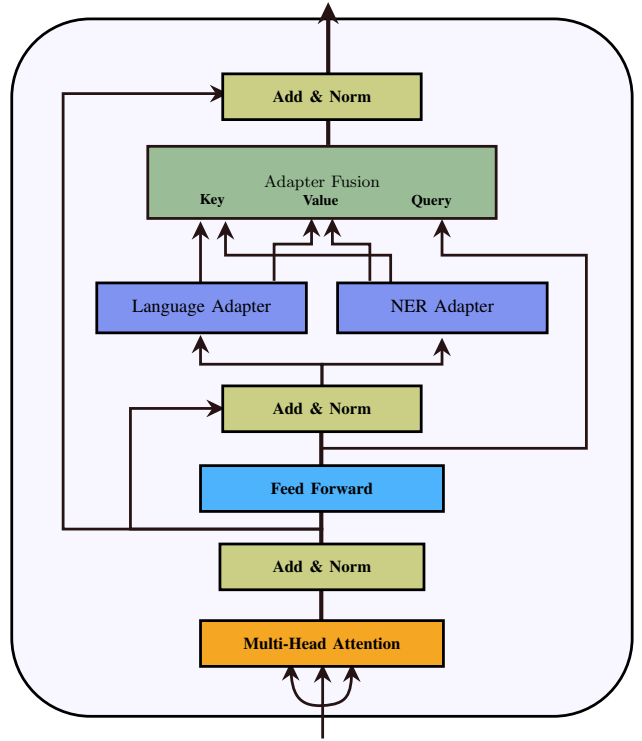
\begin{figure}
\centering
    \scalebox{.7}{
    \tikzset{every picture/.style={line width=0.75pt}} 

\begin{tikzpicture}[x=0.75pt,y=0.75pt,yscale=-1,xscale=1]

\draw  [fill={rgb, 255:red, 248; green, 247; blue, 255 }  ,fill opacity=1 ][line width=1.5]  (116,74.29) .. controls (116,43) and (141.37,17.63) .. (172.66,17.63) -- (506.34,17.63) .. controls (537.63,17.63) and (563,43) .. (563,74.29) -- (563,463.84) .. controls (563,495.13) and (537.63,520.5) .. (506.34,520.5) -- (172.66,520.5) .. controls (141.37,520.5) and (116,495.13) .. (116,463.84) -- cycle ;
\draw [color={rgb, 255:red, 38; green, 19; blue, 19 }  ,draw opacity=1 ][line width=1.5]    (340,536.5) -- (340,487) ;
\draw [shift={(340,483)}, rotate = 90] [fill={rgb, 255:red, 38; green, 19; blue, 19 }  ,fill opacity=1 ][line width=0.08]  [draw opacity=0] (13.4,-6.43) -- (0,0) -- (13.4,6.44) -- (8.9,0) -- cycle    ;
\draw [color={rgb, 255:red, 38; green, 19; blue, 19 }  ,draw opacity=1 ][line width=1.5]    (317.06,485.77) .. controls (317.63,504.01) and (322.33,513.68) .. (340,512.75) .. controls (357.86,511.81) and (360.7,505.35) .. (360.98,486.3) ;
\draw [shift={(361,482.5)}, rotate = 90] [fill={rgb, 255:red, 38; green, 19; blue, 19 }  ,fill opacity=1 ][line width=0.08]  [draw opacity=0] (13.4,-6.43) -- (0,0) -- (13.4,6.44) -- (8.9,0) -- cycle    ;
\draw [shift={(317,481.5)}, rotate = 90] [fill={rgb, 255:red, 38; green, 19; blue, 19 }  ,fill opacity=1 ][line width=0.08]  [draw opacity=0] (13.4,-6.43) -- (0,0) -- (13.4,6.44) -- (8.9,0) -- cycle    ;
\draw [color={rgb, 255:red, 38; green, 19; blue, 19 }  ,draw opacity=1 ][line width=2.25]    (339,428.5) -- (339,457) ;
\draw [color={rgb, 255:red, 38; green, 19; blue, 19 }  ,draw opacity=1 ][line width=2.25]    (339,366.5) -- (339,395.5) ;
\draw [color={rgb, 255:red, 38; green, 19; blue, 19 }  ,draw opacity=1 ][line width=2.25]    (339,310.5) -- (339,344) ;
\draw [color={rgb, 255:red, 38; green, 19; blue, 19 }  ,draw opacity=1 ][line width=1.5]    (338.5,385.5) -- (201.5,385.5) -- (201.5,298.5) -- (265.5,298.5) ;
\draw [shift={(269.5,298.5)}, rotate = 180] [fill={rgb, 255:red, 38; green, 19; blue, 19 }  ,fill opacity=1 ][line width=0.08]  [draw opacity=0] (13.4,-6.43) -- (0,0) -- (13.4,6.44) -- (8.9,0) -- cycle    ;
\draw [color={rgb, 255:red, 38; green, 19; blue, 19 }  ,draw opacity=1 ][line width=1.5]    (270,167.5) -- (270,188) -- (389,188) -- (389,208) ;
\draw [shift={(270,163.5)}, rotate = 90] [fill={rgb, 255:red, 38; green, 19; blue, 19 }  ,fill opacity=1 ][line width=0.08]  [draw opacity=0] (13.4,-6.43) -- (0,0) -- (13.4,6.44) -- (8.9,0) -- cycle    ;
\draw [color={rgb, 255:red, 38; green, 19; blue, 19 }  ,draw opacity=1 ][line width=1.5]    (252,167.5) -- (252,208) ;
\draw [shift={(252,163.5)}, rotate = 90] [fill={rgb, 255:red, 38; green, 19; blue, 19 }  ,fill opacity=1 ][line width=0.08]  [draw opacity=0] (13.4,-6.43) -- (0,0) -- (13.4,6.44) -- (8.9,0) -- cycle    ;
\draw [color={rgb, 255:red, 38; green, 19; blue, 19 }  ,draw opacity=1 ][line width=2.25]    (344,90.5) -- (344,108.5) ;
\draw [color={rgb, 255:red, 38; green, 19; blue, 19 }  ,draw opacity=1 ][line width=1.5]    (340.5,385.5) -- (153,385.5) -- (153,71.5) -- (268,71.5) ;
\draw [shift={(272,71.5)}, rotate = 180] [fill={rgb, 255:red, 38; green, 19; blue, 19 }  ,fill opacity=1 ][line width=0.08]  [draw opacity=0] (13.4,-6.43) -- (0,0) -- (13.4,6.44) -- (8.9,0) -- cycle    ;
\draw [color={rgb, 255:red, 38; green, 19; blue, 19 }  ,draw opacity=1 ][line width=2.25]    (344,10.5) -- (344,57.5) ;
\draw [shift={(344,5.5)}, rotate = 90] [fill={rgb, 255:red, 38; green, 19; blue, 19 }  ,fill opacity=1 ][line width=0.08]  [draw opacity=0] (16.07,-7.72) -- (0,0) -- (16.07,7.72) -- (10.67,0) -- cycle    ;
\draw [color={rgb, 255:red, 38; green, 19; blue, 19 }  ,draw opacity=1 ][line width=1.5]    (339,327.25) -- (530,327.25) -- (530,188) -- (426,188) -- (426,166) ;
\draw [shift={(426,162)}, rotate = 90] [fill={rgb, 255:red, 38; green, 19; blue, 19 }  ,fill opacity=1 ][line width=0.08]  [draw opacity=0] (13.4,-6.43) -- (0,0) -- (13.4,6.44) -- (8.9,0) -- cycle    ;
\draw [color={rgb, 255:red, 38; green, 19; blue, 19 }  ,draw opacity=1 ][line width=1.5]    (347,165.5) -- (347,180) -- (374,180) -- (374,206.5) ;
\draw [shift={(347,161.5)}, rotate = 90] [fill={rgb, 255:red, 38; green, 19; blue, 19 }  ,fill opacity=1 ][line width=0.08]  [draw opacity=0] (13.4,-6.43) -- (0,0) -- (13.4,6.44) -- (8.9,0) -- cycle    ;
\draw [color={rgb, 255:red, 38; green, 19; blue, 19 }  ,draw opacity=1 ][line width=1.5]    (332,165.5) -- (332,180) -- (305,180) -- (305,206.5) ;
\draw [shift={(332,161.5)}, rotate = 90] [fill={rgb, 255:red, 38; green, 19; blue, 19 }  ,fill opacity=1 ][line width=0.08]  [draw opacity=0] (13.4,-6.43) -- (0,0) -- (13.4,6.44) -- (8.9,0) -- cycle    ;
\draw [color={rgb, 255:red, 38; green, 19; blue, 19 }  ,draw opacity=1 ][line width=1.5]    (252,249.5) -- (252,266.63) -- (339.5,266.63) -- (339.5,285.5) ;
\draw [shift={(252,245.5)}, rotate = 90] [fill={rgb, 255:red, 38; green, 19; blue, 19 }  ,fill opacity=1 ][line width=0.08]  [draw opacity=0] (13.4,-6.43) -- (0,0) -- (13.4,6.44) -- (8.9,0) -- cycle    ;
\draw [color={rgb, 255:red, 38; green, 19; blue, 19 }  ,draw opacity=1 ][line width=1.5]    (426,251.5) -- (426,266.63) -- (339.5,266.63) -- (339.5,285.5) ;
\draw [shift={(426,247.5)}, rotate = 90] [fill={rgb, 255:red, 38; green, 19; blue, 19 }  ,fill opacity=1 ][line width=0.08]  [draw opacity=0] (13.4,-6.43) -- (0,0) -- (13.4,6.44) -- (8.9,0) -- cycle    ;

\draw  [fill={rgb, 255:red, 245; green, 166; blue, 35 }  ,fill opacity=1 ][line width=1.5]   (252,451.5) -- (427,451.5) -- (427,484.5) -- (252,484.5) -- cycle  ;
\draw (339.5,468) node   [align=left] {\begin{minipage}[lt]{116.28pt}\setlength\topsep{0pt}
\begin{center}
{\small \textbf{Multi-Head Attention}}
\end{center}

\end{minipage}};
\draw  [fill={rgb, 255:red, 202; green, 207; blue, 133 }  ,fill opacity=1 ][line width=1.5]   (266,396.25) -- (413,396.25) -- (413,429.25) -- (266,429.25) -- cycle  ;
\draw (339.5,412.75) node   [align=left] {\begin{minipage}[lt]{97.24pt}\setlength\topsep{0pt}
\begin{center}
\textbf{{\small Add \& Norm}}
\end{center}

\end{minipage}};
\draw  [fill={rgb, 255:red, 77; green, 179; blue, 252 }  ,fill opacity=1 ][line width=1.5]   (252,339.75) -- (427,339.75) -- (427,372.75) -- (252,372.75) -- cycle  ;
\draw (339.5,356.25) node   [align=left] {\begin{minipage}[lt]{116.28pt}\setlength\topsep{0pt}
\begin{center}
\textbf{{\small Feed Forward }}
\end{center}

\end{minipage}};
\draw  [color={rgb, 255:red, 38; green, 19; blue, 19 }  ,draw opacity=1 ][fill={rgb, 255:red, 154; green, 189; blue, 151 }  ,fill opacity=1 ][line width=1.5]   (214,109.5) -- (465,109.5) -- (465,161.5) -- (214,161.5) -- cycle  ;
\draw (339.5,135.5) node   [align=left] {\begin{minipage}[lt]{167.96pt}\setlength\topsep{0pt}
\begin{center}
\textbf{{\small {\fontfamily{helvet}\selectfont Adapter Fusion }}}
\end{center}

\end{minipage}};
\draw  [fill={rgb, 255:red, 202; green, 207; blue, 133 }  ,fill opacity=1 ][line width=1.5]   (268,282.25) -- (411,282.25) -- (411,315.25) -- (268,315.25) -- cycle  ;
\draw (339.5,298.75) node   [align=left] {\begin{minipage}[lt]{94.52pt}\setlength\topsep{0pt}
\begin{center}
\textbf{{\small Add \& Norm}}
\end{center}

\end{minipage}};
\draw  [fill={rgb, 255:red, 202; green, 207; blue, 133 }  ,fill opacity=1 ][line width=1.5]   (268,57.25) -- (411,57.25) -- (411,90.25) -- (268,90.25) -- cycle  ;
\draw (339.5,73.75) node   [align=left] {\begin{minipage}[lt]{94.52pt}\setlength\topsep{0pt}
\begin{center}
\textbf{{\small Add \& Norm}}
\end{center}

\end{minipage}};
\draw  [fill={rgb, 255:red, 128; green, 147; blue, 241 }  ,fill opacity=1 ][line width=1.5]   (177.25,208) -- (328.25,208) -- (328.25,244) -- (177.25,244) -- cycle  ;
\draw (252.75,226) node   [align=left] {\begin{minipage}[lt]{99.96pt}\setlength\topsep{0pt}
\begin{center}
Language Adapter
\end{center}

\end{minipage}};
\draw  [fill={rgb, 255:red, 128; green, 147; blue, 241 }  ,fill opacity=1 ][line width=1.5]   (350.75,208) -- (501.75,208) -- (501.75,244) -- (350.75,244) -- cycle  ;
\draw (426.25,226) node   [align=left] {\begin{minipage}[lt]{99.96pt}\setlength\topsep{0pt}
\begin{center}
NER Adapter
\end{center}

\end{minipage}};
\draw (260.5,148.25) node   [align=left] {\begin{minipage}[lt]{26.52pt}\setlength\topsep{0pt}
\begin{center}
\textbf{{\footnotesize Key}}
\end{center}

\end{minipage}};
\draw (338.5,148.25) node   [align=left] {\begin{minipage}[lt]{26.52pt}\setlength\topsep{0pt}
\begin{center}
\textbf{{\footnotesize Value}}
\end{center}

\end{minipage}};
\draw (418.5,149.25) node   [align=left] {\begin{minipage}[lt]{26.52pt}\setlength\topsep{0pt}
\begin{center}
\textbf{{\footnotesize Query}}
\end{center}

\end{minipage}};

\end{tikzpicture}
    }
\caption{
The proposed architecture for incorporating syntactical information into transformer blocks consists of two key components: a language adapter and a named entity recognition (NER) adapter. These adapters have been trained separately before being inserted into a parallel stack. An AdapterFusion module is placed on top of the stack, which is trained to perform a specific target task, such as code refinement or code summarization. This allows for the integration and combination of the knowledge gained from the below adapters, leading to more accurate and efficient performance.}
\label{fig:nerarch}

\end{figure}


\subsection{Token Type Classification Loss (TTC)}
\label{subsec:tokentypeloss}
Given a token type assigned to each token in the code sample, we pose the problem as a token classification problem, where the adapter is responsible for predicting the type of each token. To train the NER adapters, we employ cross-entropy loss function.

Let $X_i = {x_1,...,x_N}$ represent a sequence of token ids for sample $i$, $T = {t_1,...,t_N}$ indicate the corresponding sequence of type ids, and $Y = {y_1,...,y_N}$ be the one-hot representation of these type ids, with each element having the size of the total number of types presented in the dataset. The cross-entropy for sample $i$ is calculated as follows:

\begin{equation}
\begin{aligned}
L_{NER} = - \sum_{t=1}^{N} Y^{T}log(P_t)
\end{aligned}
\end{equation}

where $L_{NER}$ indicates the loss function and $P_t$ is the probability distribution for types we receive from the network for token $t$. This loss function allows us to measure the difference between the predicted and actual token types, providing a means to adjust the model's parameters accordingly and improve its performance.

\subsection{Training NER Adapters}
This section explains the steps we need to train the NER adapters.

\subsubsection{Extracting NERs Corresponding to The Leaf Nodes} 
NER adapter requires to be trained on labeled data of token types. To accurately extract the type of each token in a code sample, we utilize the tree-sitter parser\footnote{https://tree-sitter.github.io/tree-sitter/}. This parser is specifically designed for language analysis and can determine the type of each word in multiple programming languages. Tree-sitter is used in previous studies, such as in GraphCodeBERT \cite{guo2020graphcodebert} to extract dataflow from the AST and in static analysis experiments on Github \cite{clem2021static}.
An example of an AST generated for a given code snippet is shown in Fig. \ref{fig:ast-example}. To obtain the token types, we feed each code snippet to the tree-sitter and then traverse the AST to extract the type of each code token. 
    
\subsubsection{Encoding Tokens by Tokenizer and Assigning Each Sub-Token The Same Token Type}
Using the steps mentioned earlier, we extract the code token types. 
However, due to the nature of tokenization, where a single word may be split into multiple tokens, a one-to-one relationship between words and tokens \textit{cannot} be assumed. Therefore, we must establish mappings between the words and their corresponding tokens and subsequently assign the same type to each token. For example, the function name 
 \verb|find\_bad\_files| would be marked as an \textit{identifier} type by tree-sitter. When tokenized, this name may be split into three tokens: ``find", ``bad," and ``files." In this case, we would mark all of them as \textit{identifier} types.
After this process, we have a list of the tokens and their corresponding types for each sample, providing a solid foundation for further analysis and processing.

\subsubsection{Fine-Tuning NER Adapters by Token Type Classification Loss Function}
The final step is to incorporate NER adapters and fine-tune their weights using the token type classification loss function. This is done while keeping the weights of the pre-trained model fixed, allowing the NER adapters to specialize in recognizing specific named entities while leveraging the general language understanding of the pre-trained model. This approach allows for improved performance in identifying named entities in the text while maintaining the model's overall accuracy.

\subsection{Architecture Overview}


The proposed architecture for incorporating syntactical information into transformer blocks is demonstrated in Fig. \ref{fig:nerarch}. 
This architecture consists of two key components: a language adapter and a named entity recognition (NER) adapter. We train these adapters separately before inserting them into a parallel stack, as shown in the figure. 
The language adapter is trained on mask language modeling loss function. A language adapter is used to learn general language knowledge that underlies the dataset. Finally, we utilize AdapterFusion to combine the knowledge gained from the language and NER adapters to provide a more robust output representation at each transformer block. 
The AdapterFusion module is placed on top of the stack and is trained to perform a specific target task. These tasks in our experiments are code refinement and code summarization. Using AdapterFusion allows for the integration and combination of the knowledge gained from the below adapters, leading to more accurate and efficient performance.

\begin{figure}

\includegraphics[width=0.5\columnwidth]{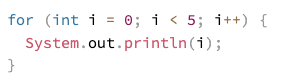}
\centering
\caption{An example of java code snippet that is sent to tree-sitter to extract its corresponding AST.}
\label{fig:code-snippet}
\end{figure}

\begin{figure}
\centering
    \scalebox{.5}{
    \input{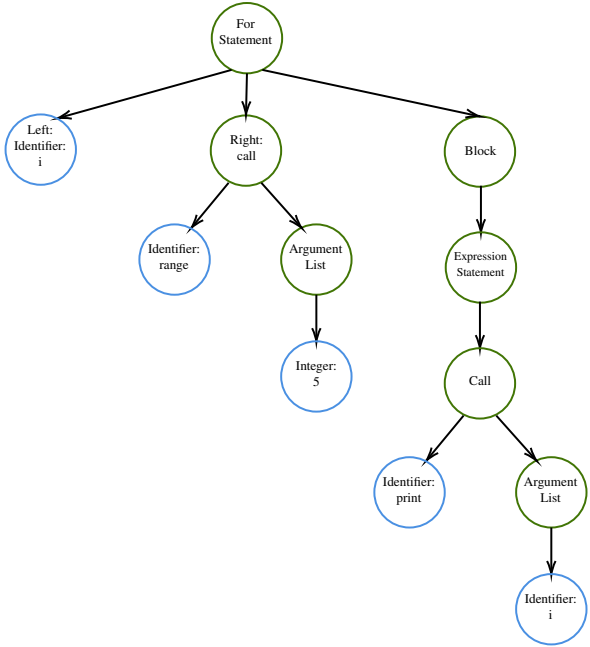}
    }
\caption{An example of an AST generated by the tree-sitter parser for input code of Fig. \ref{fig:code-snippet}. The leaf nodes are represented by blue circles, while the non-leaf nodes are represented by green circles. This AST is produced when the code snippet is fed into the parser.}
\label{fig:ast-example}

\end{figure}

\begin{figure*}
\centering
    \begin{adjustbox}{width=\textwidth}
    \input{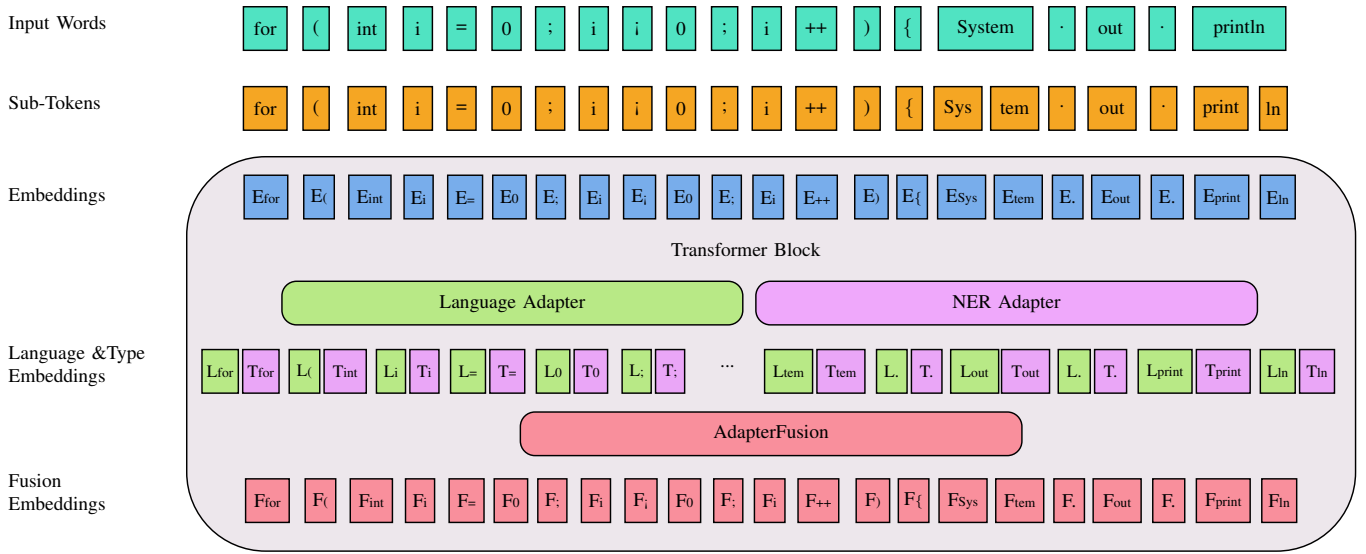}
    \end{adjustbox}

\caption{The input data flow for the sample shown in Figure \ref{fig:code-snippet} proceeds as follows when fed into a transformer block equipped with NER and language adapters. Prior to input embeddings, words are divided into sub-tokens as necessary. These sub-token embeddings, denoted by $E_{tokens}$, are then passed through both NER and language adapters in parallel. This results in two embeddings for each sub-token, denoted by $T_{token}$ and $L_{token}$, corresponding to the NER and language adapters, respectively. Finally, these embeddings are composed by AdapterFusion to select the useful information from the previous embeddings for a specific target task, such as code refinement. The final embeddings are denoted by $F_{token}$.}
\label{fig:ner-embeddings}

\end{figure*}

The details of the input data and how the embeddings of sub-tokens and adapters are composed in the AdapterFusion are shown through an example in Fig. \ref{fig:ner-embeddings}. The input code sample is presented in Fig. \ref{fig:code-snippet}, and Fig. \ref{fig:ner-embeddings} demonstrates the details. 
The code sample is fed into a transformer block equipped with NER and language adapters. Prior to input embeddings, words are divided into sub-tokens as necessary. These sub-token embeddings, denoted by $E_{tokens}$, are then passed through both NER and language adapters in parallel. This results in two embeddings for each sub-token, denoted by $T_{token}$ and $L_{token}$, corresponding to the NER and language adapters, respectively. Finally, these embeddings are composed by AdapterFusion to select the useful information from the previous embeddings for a specific target task, such as code refinement. The final embeddings are denoted by $F_{token}$.



\section{Experiment Setup}
\label{sec:expset}

We conduct experiments using CodeBERT as our backbone model for code refinement and code summarization. The former task aims to identify and fix bugs automatically, and the latter refers to automatically generating descriptions of the functionality of a given code snippet in natural language. CodeBERT is chosen as it has been studied and evaluated in several software engineering works previously \cite{feng2020codebert,tang2020multilingual,lu2021codexglue}.
Code refinement is chosen as it is heavily reliant on syntactical information to identify and correct errors and optimize code structure, and code summarization is chosen as it allows us to assess the models' generation ability and evaluate the semantic knowledge that underlies the programming and natural language aspects of a PPLM.
In the following, we explain the downstream tasks and the datasets we used to perform experiments, the baseline model, and the evaluation metrics for each downstream task.


\subsection{Downstream Tasks and Datasets}
\label{subsec:target-task}
\textbf{Code Refinement} is an essential aspect of software development and a crucial step in ensuring the robustness and reliability of software systems \cite{guo2020graphcodebert}. By leveraging various techniques and tools, it aims to automatically identify and fix bugs in the code, thus significantly reducing the cost and effort associated with resolving them manually. In our study, we utilize the dataset released by Tufano et al. \cite{tufano2019empirical} to evaluate the effectiveness of our proposed architecture and NER adapters on code refinement. We performed our experiments on BFP small dataset. The total number of datapoints in training, test, and validation sets are $46680$,
$5835$ and $5835$, respectively.

\textbf{Source Code Summarization} is a powerful technique in software engineering that aims to provide a natural language description of the source code's functionality, thus facilitating the understanding and maintenance of the code \cite{hu2018summarizing}. The goal of source code summarization is to make it easier to comprehend the overall structure and functionality of the code.
This task is widely studied, and various approaches are proposed, particularly utilizing programming language models such as CodeBERT \cite{feng2020codebert}, CodeT5 \cite{wang2021codet5}, and Multilingual CodeBERT \cite{ahmed2021multilingual} to summarize source code automatically.

We choose code refinement as it is a more abstract and complex task than type inference when we have NER adapters. More specifically, we provided AST information to evaluate the model on more complex tasks rather than type inference. We selected code summarization as it is a generative task, and the goal is to assess to what extent our proposed structural information could be helpful for NL-PL generative target tasks.

Our research uses CodeSearchNet \cite{husain2019codesearchnet} as our dataset for training the language and NER adapters. 
The dataset consists of six programming languages. The size of each language is demonstrated in Table \ref{table:plm}. 
{As tree-sitter treats the entire PHP code snippet as a single text element and fails to provide syntactical information, we exclude PHP from our experiments. Instead, we will focus on utilizing the tree-sitter for the following languages: Go, Java, JavaScript, Python, and Ruby.}
More specifically, for code refinement, language and NER adapters are trained on the training split of the unimodal Java dataset in CodeSearchNet, and AdapterFusion is trained on the Java bug fix dataset introduced in \cite{tufano2019empirical}. For code summarization, language and NER adapters are trained on the training split of the unimodal data for all languages (combined) available in CodeSearchNet. The AdapterFusion is then trained on the aggregated training splits of the bimodal data (i.e., code and comments) for all languages included in CodeSearchNet, allowing for a comprehensive and \textit{multilingual} approach to code summarization. 

NER and language adapters are trained for 50,000 training steps with a batch size of 48 and a learning rate of $2e-5$. For code refinement, AdapterFusion is trained for 100,000 training steps with a batch size of 16 and a learning rate of $5e-5$. For code summarization, we trained AdapterFusion for 50,000 training steps with a batch size of 32.






\subsection{Baseline Model}
\label{subsec:baselines}
Our approach is designed to be modular and flexible, allowing for easy integration of different pre-trained programming language models. This allows for easy experimentation and comparison of different models and their performance on various tasks.

In our study, we choose CodeBERT \cite{feng2020codebert} as our backbone and baseline model. This model is pre-trained on the CodeSearchNet dataset \cite{husain2019codesearchnet} and is widely studied in software engineering. CodeBERT is a multilingual model that is fine-tuned on monolingual datasets.

To evaluate the effectiveness of our approach, we compare the results of fine-tuning our proposed architecture when it is inserted in CodeBERT with the results of fine-tuning CodeBERT on the two tasks. This allows us to measure the potential improvement in the performance and its effectiveness when applied to code refinement and summarization tasks. Additionally, this comparison provides insight into the strengths and limitations of our proposed approach and how it compares to the baseline model.
For each task, other methods/models are used for comparison, which are explained in the Results section to save space and reduce redundancy.

\begin{table}[h!]
\centering
\caption{Statistics of the CodeSearchNet dataset for the five languages in our experiments \cite{husain2019codesearchnet}}
\begin{tabular}{|c | c | c|} 
 \hline
 \textbf{Language} & \textbf{Bimodal Data} & \textbf{Unimodal Data} \\ [0.5ex] 
 \hline
 Go  & 317,832 & 726,768  \\ 
 Java  & 500,754 & 1,569,889 \\ 
 JavaScript  & 143,252 & 1,857,835 \\ 
 Python  & 458,219 & 1,156,085 \\ 
 Ruby  & 52,905 & 164,048 \\ 
 \hline
\end{tabular}

\label{table:plm}
\end{table}

\subsection{Evaluation Metrics}

The code refinement task is evaluated using a combination of BLEU-4 score and accuracy measurements. The BLEU score, a widely accepted evaluation metric in natural language processing, is utilized to gauge the similarity between the generated output and the correct answer. Additionally, we measure accuracy, which represents the proportion of samples that the model was able to fix correctly. 
For code refinement, we follow the CodeXGLUE and report both BLEU and accuracy scores to reflect the model's performance, as considering only one of these scores can be misleading. For example, the BLEU score of the ``naive copy" is more than the BLEU score of all the baselines. The ultimate goal of our evaluation process is to optimize both of these metrics simultaneously, striving for high levels of both similarity and correction accuracy.

The code summarization task is evaluated using
BLEU (i.e., Bilingual Evaluation Understudy) score, a precision-based measurement for evaluating the performance of NLP models, particularly machine translation systems. It works by comparing the output of the language models (e.g., a generated summary) with a reference summary produced by a human expert and calculating the degree of overlap between the two. In this work we calculate BLUE-4 score \cite{lin2004orange}
in which the algorithm first identifies the unigrams to 4-grams in both the output and the reference translation and then calculates the precision between the generated summary (i.e., n-gram hit)
and the ground truth summary (i.e., total n-gram count). Higher BLEU scores indicate that the model's output is more similar to the reference translation and is therefore considered higher quality.
BLEU is commonly used in code summarization to evaluate and compare the quality and coherence of the generated summaries from different PPLMs \cite{hu2018summarizing,wang2021codet5,guo2020graphcodebert,ahmed2022learning}.


\begin{table*}
\centering
\caption{The table presents the evaluation metrics, including Precision, Recall, F1\_Score, and Accuracy, for the training phase of NER adapters. The programming languages included in the table are organized in order of resource availability, with those with the least resources available (i.e., size of the dataset) listed first and those with the most resources listed last. These metrics provide a comprehensive assessment of the performance of the NER adapters.}
\begin{tabular}{|l | c | c| c| c|} 
\hline
 \textbf{Languages} & \textbf{Precision} & \textbf{Recall} & \textbf{F1\_Score} & \textbf{Accuracy } \\ [0.5ex] 
 \hline
 
 Ruby  & 0.68 & 0.65 & 0.66 & 0.92  \\ 
 JavaScript  & 0.78 & 0.79 &0.78 & 0.91 \\ 
 Go  & 0.78 & 0.82  &0.80 & 0.94 \\ 
 Python  & \textbf{0.95} & \textbf{0.94} &\textbf{0.94} & \textbf{0.98} \\ 
 Java   & 0.78 & 0.79 & 0.78 & 0.89 \\ 
 \hline
\end{tabular}

\label{table:ner-accuracies}
\end{table*}

\section{Results and Discussion}
\label{sec:results}

\subsection{Performance of NER Adapters}

Table \ref{table:ner-accuracies} presents the evaluation metrics of the token type classification loss function, namely, Precision, Recall, F1\_score, and Accuracy, during the NER adapter training phase. 
Precision is a metric that measures the proportion of true positive predictions out of all positive predictions. Recall measures the proportion of true positive predictions from all actual positive observations. F1\_Score is the harmonic mean of Precision and Recall, and 
Accuracy is a metric that measures the proportion of correct predictions out of all observations. These NER adapters were trained on the types obtained from the tree-sitter parser on the CodeSearchNet dataset. 

The programming languages in the table are arranged in order of dataset size, with the least resource-intensive languages listed first and the most resource-intensive languages listed last. 
Python achieves the best performance for all metrics compared to other languages, and Ruby has the least performance as it suffers from a low resource dataset. 
It is worth noting that the precision, recall, and F1\_scores are highly dependent on the size of the dataset, indicating that larger datasets generally yield better results for NER adapter training. 
However, the results for Java are worse than those of other languages, despite having a relatively rich dataset. This discrepancy may be attributed to the syntactical complexities present in Java compared to Python, which contributed to its lower performance. 
On the other hand, Python exhibits exceptional performance across all the evaluated metrics. The high levels of accuracy recorded for all programming languages further demonstrate that the NER adapters are effectively trained to extract the syntactical information from the network's internal embeddings. This highlights the efficiency and robustness of our NER adapter implementation.


\subsection{CodeBERTER' Results for Code Refinement}

We evaluate the performance of NER adapters on code refinement task to check to what extent they could be effective on a target programming task. The results are presented in Table \ref{table:code-refinement}. 
The \textbf{Naive copy} adopts a simplistic approach by directly copying the buggy code as the refinement result. 
The \textbf{LSTM} use the Long Short Term Memory architecture, and the \textbf{Transformer} model utilizes 12 transformer encoder blocks (i.e., the same number of layers and hidden size as pre-trained models). 
For the Transformer, the encoder is initialized with pre-trained models, while the decoder's parameters are randomly initialized. The results depicted in the table demonstrate that the Transformer significantly outperforms the LSTM model. 

We separate the results of these models in Table \ref{table:code-refinement} from the second group of models that leverage pre-training on programming languages. 
The BLEU and Accuracy scores are higher for the models with pre-training.
\textbf{RoBERTa (code)} is pre-trained only on code. \textbf{NSEdit} \cite{hu2022fix} propose a new pointer network to transformer blocks to predict edit locations. \textbf{CoTexT} \cite{phan2021cotext} is a pre-trained bimodal encoder-decoder programming model.
Among the programming pre-trained models, our approach, denoted by \textbf{CodeBERTER}, stands out by achieving the best performance on BLEU score. Note that \textbf{CodeBERTER} improves the results of CodeBERT \textit{without} taking advantage of fine-tuning all of its parameters. This illustrates the beneficial impact of incorporating code structure information in code refinement.

\begin{table}[t!]
\centering
\caption{NER adapter BLEU score and Accuracy results on code refinement task. The baselines are divided by a horizontal line, where the bottom group shows the models that are pre-trained on programming languages. The bold font indicates the best results, which belongs to CodeBERTER.}
\begin{tabular}{|l | c | c|} 
 \hline
 \textbf{Method/Model} & \textbf{BLEU} & \textbf{Accuracy}  \\ [0.5ex] 
 \hline
 Naive copy   & 78.06 & 0.0    \\ 
 LSTM  & 76.76 & 10.0  \\ 
 Transformer  & 77.21 & 14.7  \\ 
 \hline
 RoBERTa (code)   & 77.30 & 15.9   \\ 
 CodeBERT  & 77.42 & 16.4  \\ 
 \textbf{CodeBERTER} & \textbf{78.2} &17.8  \\ 
CoTexT  & 77.91 & 22.64  \\ 
NSEdit  & 71.06 & \textbf{24.04}  \\ 
 \hline
\end{tabular}

\label{table:code-refinement}
\end{table}

\begin{table*}[t!]
\centering
\caption{Smooth BLEU-4 scores on code summarization. CodeBERTER is fine-tuned on the multilingual datasets (same as $polyglot$CodeBERT). The horizontal line separates our approach from the others, and the dashed-horizontal line separates the language models fine-tuned on multilingual datasets from the ones fine-tuned on monolingual datasets. For readability, the results of CodeBERT are underlined. The best scores among all models for each language are shown in bold.}
\begin{tabular}{|l | c | c| c| c| c| c|} 
 \hline
 \textbf{Models} & \textbf{Ruby} & \textbf{JavaScript} & \textbf{Go} & \textbf{Python} & \textbf{Java} & \textbf{Average} \\ [0.5ex] 
 \hline
 CodeBERTER & \textbf{15.90} & \textbf{16.12} & \textbf{23.34} & 18.38 & 19.95 & \textbf{18.738}  \\ 

\hline
 
 $polyglot$GraphCodeBERT \cite{ahmed2021multilingual} & 14.95 & 15.79 &18.92 & 18.90 & 19.91 & 17.694 \\ 
 $polyglot$CodeBERT \cite{ahmed2021multilingual} & 14.75 & 15.80 &18.77 & 18.71 & 20.11 & 17.48 \\ 
 \hdashline
 DistillCodeT5 & 15.75 & 16.42 &20.21 & \textbf{20.59} & \textbf{20.51} & 	18.696 \\ 
 CodeT5 \cite{wang2021codet5} & 15.69 & 16.24 &19.76 & 20.36 & 20.46 & 18.502 \\ 
 ProphetNet-Code \cite{qi2021prophetnet} & 14.37 & 16.60  &18.43 & 17.87 & 19.39 & 	17.332 \\ 
CoTexT \cite{qi2021prophetnet} & 14.02 & 14.96  &18.86  & 19.73 & 	19.06 & 	17.326 \\ 
 PLBART \cite{ahmad2021unified} & 14.11 & 15.56 &18.91 & 19.30 & 18.45 & 17.22 \\ 
 GraphCodeBERT   & 12.62 & 14.79 &18.40 & 18.02 & 19.22 & 16.61 \\ 
 CodeBERT  & \underline{12.16} & \underline{14.90} & \underline{18.07} & \underline{19.06} & \underline{17.65} & 16.36 \\ 
 RoBERTa \cite{zhuang2021robustly}  & 11.17 & 11.90 &17.72 & 18.14 & 16.47  & 15.08 \\ 
 Transformer \cite{vaswani2017attention} & 11.18 & 11.59 &16.38 & 15.81 & 16.26 & 14.24\\ 
 seq2seq \cite{sutskever2014sequence}  & 9.64 & 10.21 &13.98 & 15.93 & 15.09 & 12.97 \\ 
 \hline
\end{tabular}

\label{table:code-summarization}
\end{table*}

\subsection{CodeBERTER's Results for Code Summarization}

Code summarization evaluates a programming language model's semantic and syntactic aspects. Table \ref{table:code-summarization} displays the results of our approach compared to other models. 
For this target task, we select a multilingual fine-tuning paradigm based on the recent findings and recommendations in \cite{ahmed2021multilingual}, where the authors mention that code summarization can benefit from multilingual fine-tuning. It means that, in our approach, first, we trained the language and NER adapters on the multilingual dataset, as mentioned earlier. We also trained the AdapterFusion on the multilingual code summarization dataset and tested it on each language separately.
Notice that for code refinement, as there exists only a Java dataset, to the best of our knowledge, we could only evaluate the performance of NER adapters on code refinement in a monolingual setting. 

Table \ref{table:code-summarization} shows the BLUE-4 scores of the state-of-the-art approaches and our approach, denoted by \textbf{CodeBERTER}, separated by a horizontal line. 
The best scores are shown in bold. 
The scores obtained from CodeBERT are also underlined for readability. CodeBERT is the primary baseline we need to compare our results with since our backbone model is CodeBERT. 
\textbf{CodeBERTER} improves the scores for four languages, Ruby, JavaScript, Go, and Java, and has on-par results with the best models for Python. 
Looking into the statistics of the data provided in Table \ref{table:plm} indicates that the first three languages have a lower number of data for training compared to the other two. For these low-resource languages, our model significantly improves the results compared to its baseline, CodeBERT. 

$polyglot$CodeBERT and $polyglot$GraphCodeBERT, as described in \cite{ahmed2021multilingual} by Ahmed et al. (2021), are models that have been fully fine-tuned in a multilingual setting. For example, to evaluate their performance on Ruby, the models were fine-tuned on all of the programming languages in the CodeSearchNet dataset \cite{husain2019codesearchnet} and subsequently evaluated on a Ruby test dataset. 
This approach has improved the performance of $polyglot$ models, as demonstrated by the difference in performance between CodeBERT and $polyglot$CodeBERT. 
However, \textbf{CodeBERTER} additionally improves the scores of $polyglot$CodeBERT, which is attributed to incorporating multilingual syntactical information into the model using NER adapters. 

For the high resource languages, Python and Java, \textbf{CodeBERTER} performs on par with the baseline model (improves Java results from CodeBERT by 2 BLEU scores). 
We relate this to the fact that for high-resource languages, the language model is able to learn the general knowledge underlying a language better than a low-resource language, for which less training data exists \cite{chen2022transferability}. 
Therefore, the NER adapter enables the model to provide more type information for low-resource languages. 


Although \textbf{CodeBERTER} has the best scores for some languages, note that \textbf{CodeBERTER} uses CodeBERT as its backbone, so it would be fairer to compare the effectiveness of our model with CodeBERT and not larger models such as CodeT5. These larger models have two times more parameters than CodeBERT. However, we considered the larger models in Table \ref{table:code-summarization} to demonstrate that our approach could even have on-par results with larger models for each of the languages. Following the leaderboard on CodeXGLUE, we also report the \textit{average} scores among the models in Table \ref{table:code-summarization}. If we compute the averages over the studied languages for code summarization, the \textbf{CodeBERTER} performance is $18.738$, which is \textbf{higher} than all other models on the leaderboard for this task.

\begin{figure}

\includegraphics[width=\columnwidth]{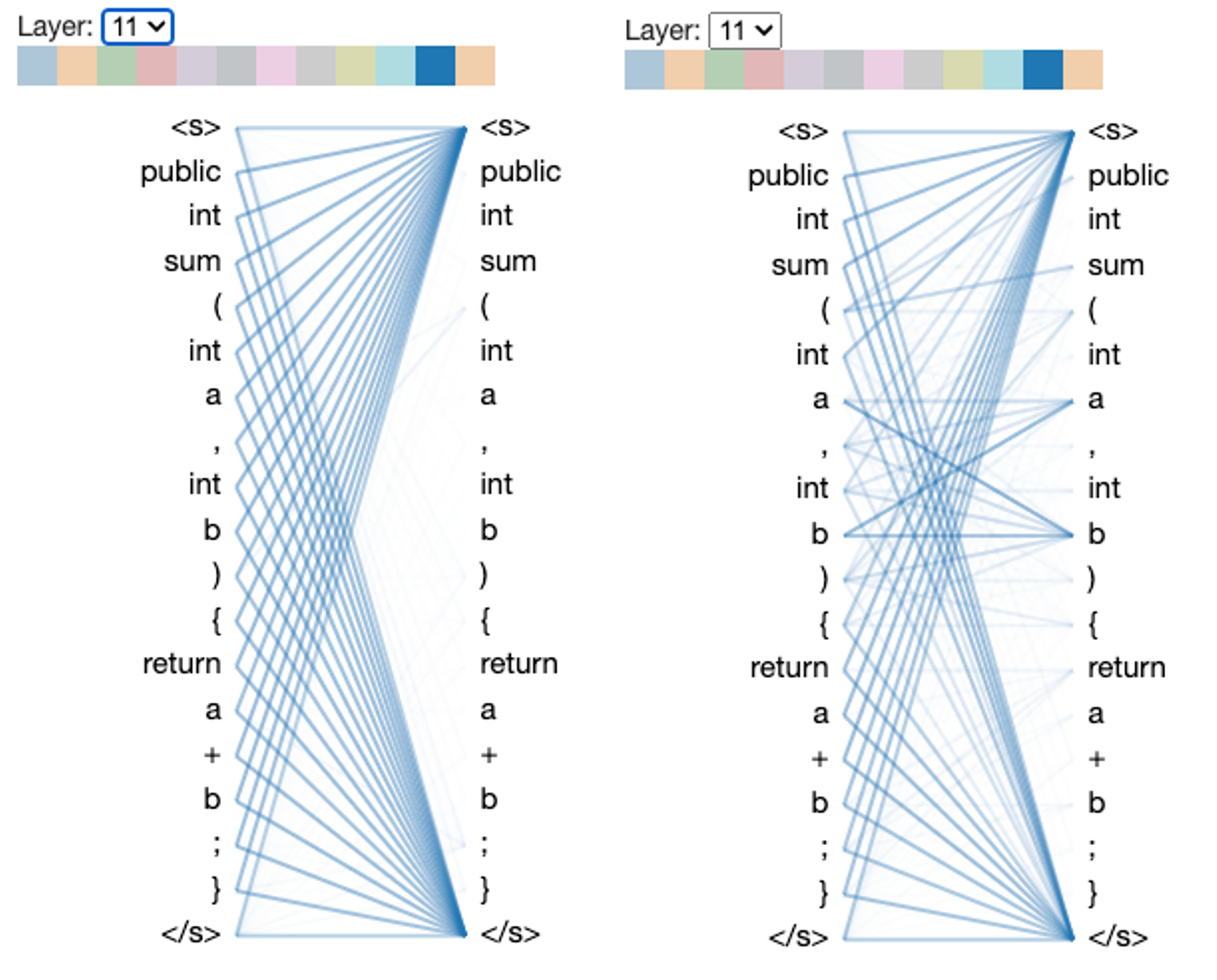}
\centering
\caption{The figure illustrates the attention patterns of the final layer of CodeBERT for a Java sample, with (the figure on the right) and without (the left figure) the insertion of NER adapters to the pre-trained model. This comparison highlights the impact of NER adapters on the model's ability to identify and extract relevant entities within the sample text.}
\label{fig:ner-effect}
\end{figure}

\subsection{Discussions}

\textbf{Computational efficiency of our approach.}
For all languages, we train NER adapters more efficiently with lower computational costs. Each of the language and NER adapters have $\sim0.9$ million and AdapterFusion has $\sim21$ trainable parameters. We train these tree adapters for code refinement in total, leading to $\sim23$ million trainable parameters. This number is still less than the total trainable parameters of the CodeBERT model, which is $\sim 110$ million. Note that all the 110 million parameters in CodeBERT should be re-trained during the standard fully fine-tuning phase.
However, this is not required in our approach. In CodeBERTER, we trained $\sim23$ million parameters for all adapters, including language, NER, and AdapterFusion. 
In terms of time efficiency, each NER adapter takes about $10$ hours to be trained, for $40,000$ training steps. This is while if we fully fine-tune CodeBERT for the NER task with the same amount of training steps, it takes around $17$ hours.


\textbf{Effect of NER adapter.}
Fig. \ref{fig:ner-effect} demonstrates the attention patterns of different tokens in a code sample on each other in the final layer of CodeBERT, generated by bertviz\footnote{https://github.com/jessevig/bertviz}. The strength of the lines among the tokens demonstrates the amount of attention; the thicker the line is, the more attention is put on that token. 
We only show one attention head (out of 12 heads) of the
last layer for preventing a messy representation in this figure.
Fig. \ref{fig:ner-effect} compares the attention patterns for a Java sample, for CodeBERT on the left, and for CodeBERTER (i.e., CodeBERT with NER adapters) on the right.
As this is shown, the utilization of NER adapters results in a shift in the attention pattern, meaning that it is more distributed among the tokens. Specifically, in CodeBERTER, the NER adapter places a greater emphasis on identifier tokens such as \verb|sum|, \verb|a|, and \verb|b|, compared to the baseline model. This allows the model to leverage the additional attention paid to these tokens for downstream tasks, resulting in improved performance.

Note that the CodeBERT attention shows multiple weak cross attentions between tokens, and most of the attention is on the $<s>$ token. This is aligned with previous findings of Sharma et al. \cite{sharma2022exploratory}, where the authors found that the special tokens (e.g., $<s>$) get the most amount of attention especially in the last layer. So, as previously shown in \cite{sharma2022exploratory}, new techniques are required to direct the BERT-based models' attention toward other tokens for programming languages. This is achieved in our work using the NER adapters.




\section{Related Work}
\label{section:related-work}

In recent years, there have been many studies focusing on code representation learning for various software engineering tasks such as code generation \cite{zeng2022extensive,zhou2022doccoder,fried2022incoder}, code summarization \cite{gu2022assemble,ahmed2022learning,nie2022impact}, program synthesis \cite{vaithilingam2022expectation,nijkamp2022conversational,ellis2021dreamcoder,austin2021program}, code search \cite{nadeem2022codedsi}, and bug repair \cite{2022arXiv220700301R,2022arXiv220805446Z}. With the advent of pre-trained language models in NLP, researchers in software engineering have leveraged the advancements in pre-trained language models in NLP to propose programming pre-trained language models that are pre-trained on either unimodal (i.e., code) or bimodal (i.e., code and comment) datasets. 

CodeBERT \cite{feng2020codebert}, GraphCodeBERT \cite{guo2020graphcodebert} and CodeT5 \cite{wang2021codet5} are the examples of this approach. Even though source code contains rich syntactical and semantical information, most of the PPLMs treat code as a sequence of tokens. GraphCodeBERT \cite{guo2020graphcodebert} extends the input with the dataflow extracted from the AST. This allows the model to better understand the relationships and dependencies between different parts of the code. TreeBERT \cite{jiang2021treebert} represents the input as a concatenation path of the AST leaves corresponding to each code sample. This allows the model to better capture the structure and hierarchy of the code. CodeT5 \cite{wang2021codet5} is a code representation learning model that is based on the T5 architecture. The model is pre-trained on a large code corpus to predict the next token in a code sequence. To improve the model's understanding of code, a new objective function was proposed in CodeT5 to inform the model about the presence of identifier tokens during the pre-training phase. The model was evaluated on several code-related tasks such as code generation, code summarization, and code search and showed improved performance compared to other existing models.

Adapters have been widely used in NLP \cite{houlsby2019parameter,pfeiffer2020mad,pfeiffer2020adapterfusion} as a quick and parameter-efficient fine-tuning approach. Adapters were previously used in software engineering to show the transferability of natural language PLMs to programming languages with lower training costs. The authors trained language adapters (i.e., adapters trained on masked language modeling on unlabelled data) and task adapters (i.e., adapters trained for a target task on labeled data) for code clone detection \cite{goel2022cross}.


\textbf{Differences of current studies with our work.} Although syntactical information is used in some of the current PPLMs, none of the models can integrate this information into an existing model without requiring pre-training it from scratch. On the other hand, NER adapters enable imposing the syntactical information in the current models without pre-training. Moreover, though adapters are widely used in NLP and are recently studied in software engineering, no work uses adapters in such a context as we do in our work. Our approach is novel not only for the TTC loss function that we introduce for NER adapters but also for how they are used (i.e., the proposed architecture in Fig. \ref{fig:nerarch}).

\section{Threats to Validity}
\label{sec:threat}

\textbf{External Validity} In this study, we evaluate the results of imposing syntactical information for code summarization on the CodeSearchNet dataset and code refinement for Java language. The task and the programming languages are restricted, and the results might not be generalizable to all downstream tasks and other datasets. 
However, based on our observations and the fact that we provide language and structural information through adapters, we hypothesize that the results for other tasks would be at least on par with the current models, emphasizing that this could be achieved with \textit{less} trainable parameters. Though still experiments should confirm this. 
Note that our approach is not restricted to CodeBERT and NER adapters can be used in other PPLMs. 

For code refinement, there exist various approaches that have been proposed in the literature. One notable contribution is the use of beam search, which has been empirically studied by Tufano et al. \cite{tufano2019empirical}. However, as our research focuses on a different aspect of code refinement, we did not explore this particular approach in depth.

\textbf{Internal Validity} Hyperparameters can affect the fine-tuning phase of a pre-trained model, and there is no hard rule to select the best values for these parameters. Herefore, the model might be subject to being stuck in sub-optimal solutions. As Pfeiffer et al. \cite{pfeiffer2020adapterfusion} performed an extensive hyperparameter search over adapters, we considered the default settings for adapters' hyperparameters. However, it is performed in the NLP area and might not lead to the optimal state in the software engineering domain.
\section{Conclusion and Future Works}
\label{section:conclusion}
In this study, we introduced NER adapters, a novel approach for enhancing existing pre-trained models by imposing syntactical information. To evaluate their performance, we conducted experiments on two programming-related tasks: code refinement using a Java dataset in a monolingual setting and code summarization using a multilingual approach. Our results showed that CodeBERTER --CodeBERT with NER adapters--outperforms the baseline models in both tasks. 
The NER adapters are model-agnostic and require fewer parameters to be trained compared to fully pre-train or fine-tuning a model.
We plan to apply CodeBERTER to other downstream tasks and NER adapters to other pre-trained models. 
\ifCLASSOPTIONcompsoc
  \section*{Acknowledgments}
\else
  \section*{Acknowledgment}
\fi

This research is supported by a grant from the Natural Sciences and Engineering Research Council of Canada RGPIN-2019-05175. 

\balance
\bibliographystyle{IEEEtran}
\bibliography{bibliography}

\begin{IEEEbiography}{Iman Saberi Tirani}
Biography text here.
\end{IEEEbiography}

\begin{IEEEbiographynophoto}{Fuxiang Chen}
Biography text here.
\end{IEEEbiographynophoto}

\begin{IEEEbiographynophoto}{Fatemeh Fard}
Biography text here.
\end{IEEEbiographynophoto}

\end{document}